# Ultrahigh Responsivity Photodetectors of Two-dimensional Covalent Organic Frameworks Integrated on Graphene

## Authors

Yi-feng Xiong[1†], Qiao-bo Liao[2†], Zheng-ping Huang[3], Xin Huang[2], Can Ke[2], Heng-tian Zhu[1], Chen-yu Dong[1], Hao-shang Wang[1], Kai Xi[2*], Peng Zhan[3], Fei Xu[1*], Yan-qing Lu[1*]

## Affiliations

[1]National Laboratory of Solid State Microstructures, College of Engineering and Applied Sciences and Collaborative Innovation Center of Advanced Microstructures, Nanjing University, Nanjing 210093, China.
[2]School of Chemistry and Chemical Engineering, Nanjing University, Nanjing 210093, China.
[3]School of Physics and National Laboratory of Solid State Microstructures, Nanjing University, Nanjing 210093, China.
[†]These authors contributed equally to this work.
[*]e-mail: xikai@nju.edu.cn; feixu@nju.edu.com; yqlu@nju.edu.cn.

## Abstract

Two-dimensional (2D) materials exhibit superior properties in electronic and optoelectronic fields. The wide demand for high-performance optoelectronic devices promotes the exploration of diversified 2D materials. Recently, 2D covalent organic frameworks (COFs) have emerged as next-generation layered materials with predesigned π-electronic skeletons and highly ordered topological structures, which are promising for tailoring their optoelectronic properties. However, COFs are usually produced as solid powders due to anisotropic growth, making them unreliable to integrate into devices. Here, by selecting tetraphenylethylene (TPE) monomers with photoelectric activity, we designed and synthesized photosensitive 2D-COFs with highly ordered topologies and grew 2D-COFs in situ on graphene to form well-ordered COF-graphene heterostructures. Ultrasensitive photodetectors were successfully fabricated with the COF$_{ETBC-TAPT}$-graphene heterostructure and exhibited an excellent overall performance with a photoresponsivity of ~$3.2 \times 10^7$ A·W$^{-1}$ at 473 nm and a time response of ~1.14 ms. Moreover, due to the high surface area and the polarity selectivity of COFs, the photosensing properties of the photodetectors can be reversibly regulated by specific target molecules. Our research provides new strategies for building advanced functional devices with programmable material structures and diversified regulation methods, paving the way for a generation of high-performance applications in optoelectronics and many other fields.

# Introduction

Along with the discovery of graphene, many two-dimensional (2D) materials—including hexagonal boron nitride (hBN), 2D chalcogenides, 2D oxides, and so on—have been exfoliated successfully[1-3]. Atom-thin 2D materials exhibit superior properties relative to their bulk materials (such as excellent mechanical flexibility, high mobility, good chemical stability, etc.)[4,5], allowing the possibility of their potential applications in nanoscale optoelectronic fields[6-11]. In recent years, several strategies have been developed to functionalize 2D material-based optoelectronic devices, such as interface modification (reducing surface charge traps, chemical doping, alloying different materials, etc.) and device structure optimization (forming a heterojunction, applying an external electric field, efficiently encapsulating, etc.)[12-17]. The construction of 2D materials is generally inherent, challenging to program by molecular structure design.

Recently, two-dimensional covalent organic frameworks (2D-COFs) have emerged as next-generation layered materials with many outstanding properties, including high thermal stabilities, permanent porosity, large surface area, and low mass densities[18,19]. More importantly, 2D-COFs are a class of organic crystalline materials with predesigned π-electronic skeletons and highly ordered topological structures that are covalently constructed from 2D building blocks. These building blocks arrange into periodic planar networks and finally stack to form layered structures aligned with atomic precision in the vertical direction[20]. Such high crystallinity and stacking alignment endow 2D-COF ordered π-electronic systems with the ability for charge carrier transport, implying that 2D-COFs have potential for optoelectronic applications[21-31]. Therefore, the optoelectronic properties of 2D-COFs are promising for programming by selecting suitable monomer combinations among abundant 2D π-electronic building blocks.

Herein, we designed and synthesized certain 2D-COFs by selecting suitable tetraphenylethylene (TPE) monomers with photoelectric activities. This structure design endows 2D-COFs with highly ordered in-plane topologies and stacking alignment with superior optoelectronic properties. However, owing to the anisotropic growth of COFs, they are usually produced as solid powders without solubility or processability, making integration of COFs into devices difficult and impeding their further applications. To solve this problem, we grew COFs in situ on chemical vapor deposition (CVD) single-layer graphene (SLG). With the assistance of the π-electron plane of graphene, the COFs tended to stacked to form oriented 2D layered structures parallel to graphene, constituting well-ordered COF-graphene heterostructures[32-35]. Additionally, benefiting from the high carrier mobility and fast response times inside graphene, high-performance COF-graphene photodetectors were successfully fabricated. In particular, due to the donor-acceptor structure[22,23], the COF$_{ETBC-TAPT}$-graphene photodetector exhibits ultrahigh photoresponsivity of ~$3.2 \times 10^7$ A·W$^{-1}$ with a fast time response of ~1.14 ms. Moreover, owing to the high surface area and the polarity selectivity of COF materials, the photosensing properties of the photodetectors can be reversibly regulated by specific target molecules.

The molecular structural design and external regulation of COFs affords opportunities to produce high performing optoelectronics.

## Results

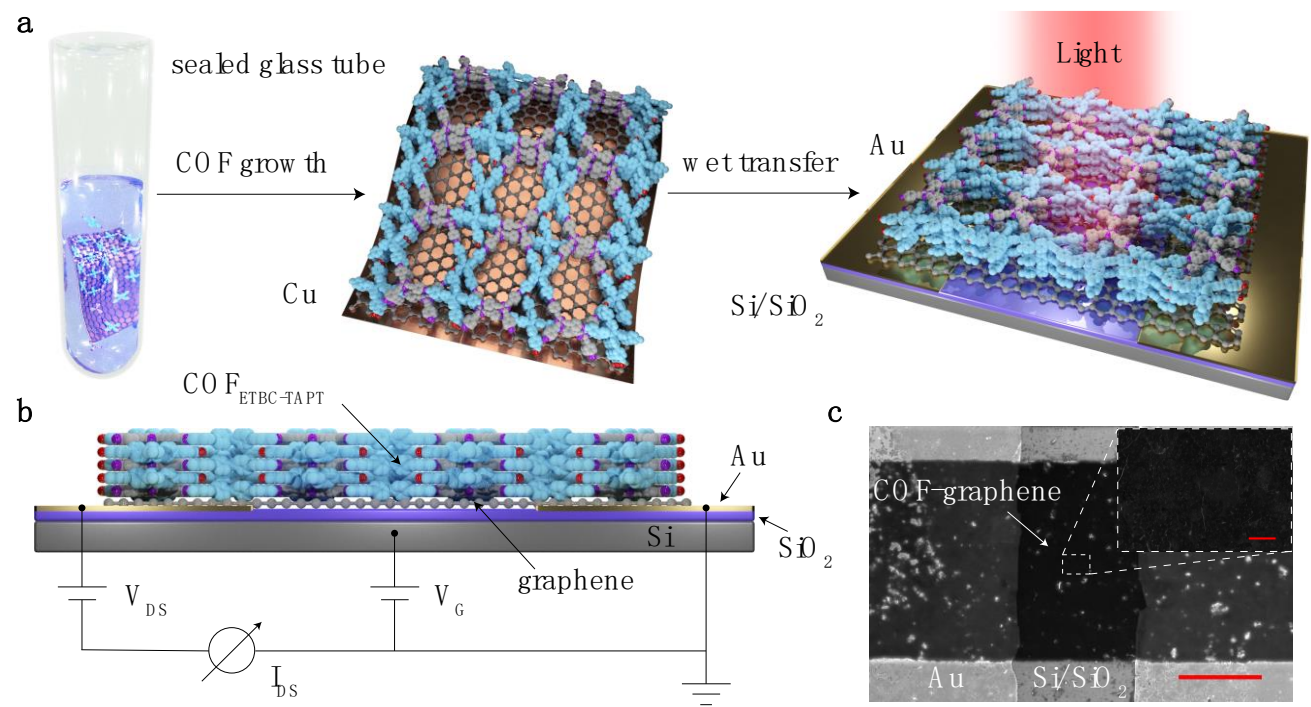

**Fig. 1. COF$_{ETBC-TAPT}$-graphene photodetectors. (a)** COF$_{ETBC-TAPT}$ was oriented grown on Cu-supported CVD graphene in a sealed glass tube. Photodetectors were fabricated by assembling a COF$_{ETBC-TAPT}$-graphene heterostructure with Au electrodes on a Si/SiO$_2$ substrate. **(b)** Side schematic view of a constructed COF$_{ETBC-TAPT}$-graphene photodetector and its measurement setup. **(c)** The SEM image of a fabricated device. The scalebar is 20 μm. Inset: The enlarged SEM image of the COF$_{ETBC-TAPT}$-graphene area. The scalebar is 1 μm.

Fig. 1a demonstrates the fabrication process of COF$_{ETBC-TAPT}$-graphene photodetector devices. Single-layer graphene (SLG) was supported by a Cu substrate submerged into a mixture of 4',4''',4'''',4'''''-(1,2-ethenediylidene)tetrakis[1,1'-biphenyl]-4-carboxaldehyde (ETBC) and 2,4,6-tris(4-aminophenyl)-1,3,5-triazine (TAPT) and co-solvent in a flame-sealed glass tube.[36,37] Under the solvothermal heating conditions, the monomers reacted with each other, and COF$_{ETBC-TAPT}$ was grown on the graphene surface to form a heterostructure. To characterize the photoelectric properties of the COF$_{ETBC-TAPT}$-graphene heterostructure, it was placed above the source-drain Au electrodes on a Si/SiO$_2$ substrate by wet transfer method, and the graphene directly contacted the Au electrodes, as shown in Fig. 1b. Photolithography and O$_2$ plasma etching were then carried out to pattern the channel (see Methods for details). The SEM images of the device are displayed in Fig. 1c, confirming a relatively clean and flat film in the channel region. Energy Dispersive X-ray (EDX) analysis (Fig. S8) ensures the purity of the COF$_{ETBC-TAPT}$-graphene heterostructure after being transferred.

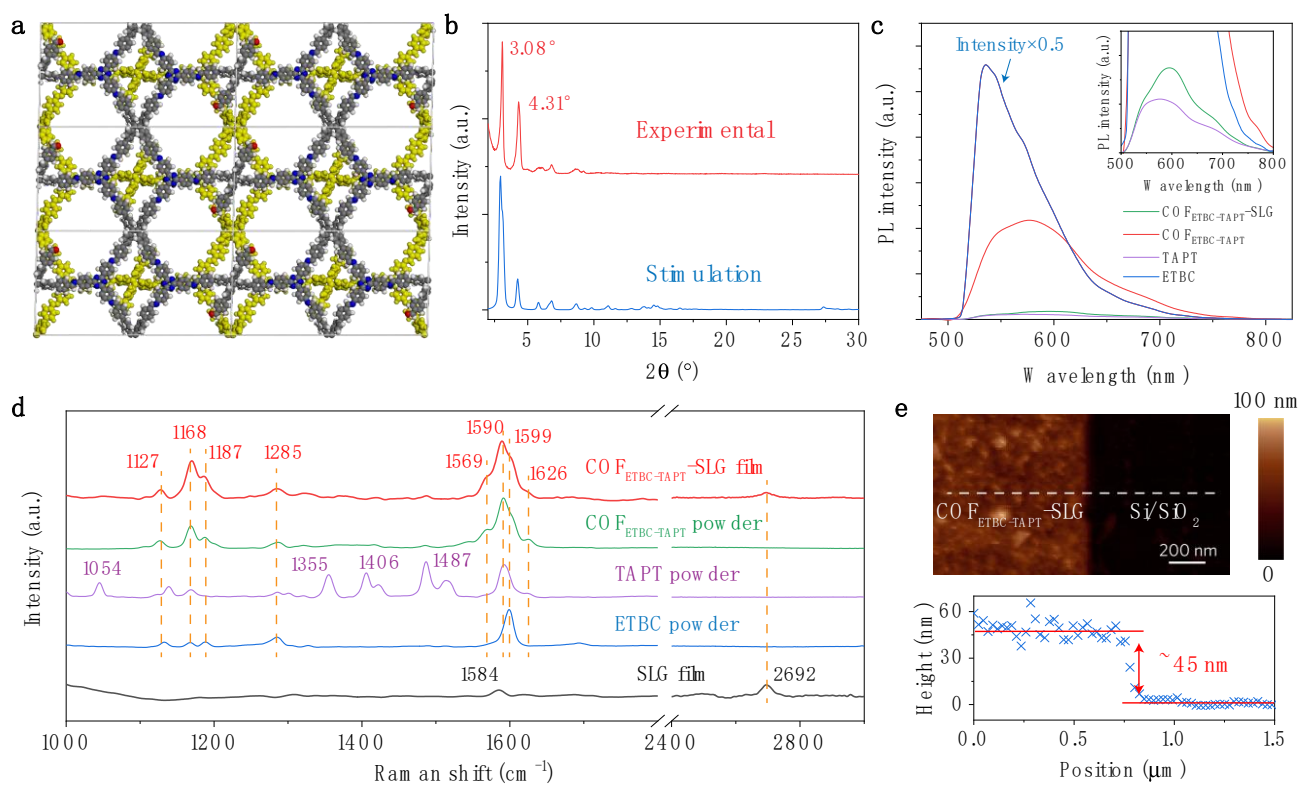

**Fig. 2. COF$_{ETBC-TAPT}$-graphene heterostructure characteristics.** (**a**) Top views of a graphical representation of a 2×3 rectangle grid showing the staggered A-B stacking of COF$_{ETBC-TAPT}$ (C, grey; N, blue; O, red; H, white and the second layer, yellow). (**b**) Comparison of the experimental PXRD pattern (top) with the simulation patterns of A-B-arranged COF$_{ETBC-TAPT}$ (bottom). (**c**) PL spectra of the COF$_{ETBC-TAPT}$-SLG film, COF$_{ETBC-TAPT}$ powder and powders of the corresponding monomers (ETBC and TAPT) on the SiO$_2$ substrate. The excitation laser is 470 nm. Inset: The enlarged image shows the PL peaks of the TAPT powder and COF$_{ETBC-TAPT}$-SLG film. (**d**) Raman spectra of the COF$_{ETBC-TAPT}$-SLG film, COF$_{ETBC-TAPT}$ powder and powders of the corresponding monomers (ETBC and TAPT) and SLG, using a 785 nm laser. (**e**) Surface topography of the COF$_{ETBC-TAPT}$-SLG film. Above: AFM topography image of the COF$_{ETBC-TAPT}$-SLG film. Bottom: Cross-section analysis of the COF$_{ETBC-TAPT}$-SLG film.

Fig. 2a demonstrates the highly ordered topological structure and staggered A-B stacking arrangement of COF$_{ETBC-TAPT}$, which can be proved by the powder X-ray diffraction (PXRD) experiments in Fig. 2b (see Fig. S2 for details). The PXRD pattern confirmed that COF$_{ETBC-TAPT}$ was highly crystalline, showing intense diffraction peaks at 3.08° and 4.31°, in accordance with the predicted PXRD pattern of the simulated structure (see modelling details in Fig. S1). The steady-state photoluminescence (PL) spectra of the COF$_{ETBC-TAPT}$-graphene film, COF$_{ETBC-TAPT}$ powder and powders of the corresponding monomers (ETBC and TAPT) were measured under excitation of 470 nm, as shown in Fig. 2c. The ETBC powder displayed a strong fluorescence emission due to the aggregation-induced emission (AIE) effect, while the TAPT powder showed a weak PL intensity. For the COF$_{ETBC-TAPT}$ powder, a wide PL peak at ~600 nm was observed, which agree well with the measured energy bandgap of COF$_{ETBC-TAPT}$

(Fig. S5). To confirm that $COF_{ETBC-TAPT}$ was successfully integrated on graphene, the PL spectrum of the $COF_{ETBC-TAPT}$-graphene film was also measured. Obviously, the PL quantum yield of the $COF_{ETBC-TAPT}$-graphene film was much lower than that of the $COF_{ETBC-TAPT}$ powder, which resulted from the effective charge carrier transfer through π–π interactions between the $COF_{ETBC-TAPT}$ layer and the graphene layer[38,39].

Then, we used Raman spectroscopy to monitor whether $COF_{ETBC-TAPT}$ was well connected with a graphene monolayer by π-π stacking. The measurements were carried out by using a commercial Renishaw confocal micro-Raman spectrometer, and all spectra were excited with a 785 nm laser and collected in the backscattering configuration. To track the origin of these characteristic Raman peaks, the Raman spectra of the monomers (*i.e.,* ETBC and TAPT powders), $COF_{ETBC-TAPT}$ powder and single-layer graphene on Cu were measured. The Raman peaks at approximately 1580 $cm^{-1}$ (G-band) and 2690 $cm^{-1}$ (2D-band) are the most notable features of single-layer graphene, shown as a black curve in Fig. 2d, and the Raman intensity of the 2D-band is almost twice that of the G-band. Compared with the Raman spectrum of the monomer TAPT powder (purple curve), the Raman peaks centred at 1054 $cm^{-1}$, 1355 $cm^{-1}$ and 1406 $cm^{-1}$ disappeared in the Raman spectrum of the $COF_{ETBC-TAPT}$ powder, which might have resulted from the π-π interaction between the monomer TAPT and ETBC powders and the polymerization of these two monomers. It is worth noting that the emerging bands of the $COF_{ETBC-TAPT}$ powder and $COF_{ETBC-TAPT}$-graphene film at 1569 $cm^{-1}$ correspond to the vibrations of newly formed imine linkages [33]. Additionally, the inter-monomer chemical interaction might strengthen the integral rigidity of the resultant COF molecule, which leads to suppression of some typical molecular vibrations of the corresponding monomer. As shown in the red curve in Fig. 2d, the as-prepared $COF_{ETBC-TAPT}$ was successfully immobilized on single-layer graphene by the chemical integration process. The surface topography of $COF_{ETBC-TAPT}$-graphene was measured by atomic force microscopy (AFM) and is displayed in Fig. 2e, which illustrates that $COF_{ETBC-TAPT}$ grew uniformly on graphene in the solvothermal reaction. By cross-section analysis of the AFM image, the thickness of the $COF_{ETBC-TAPT}$-graphene film was evaluated to be ~45 nm.

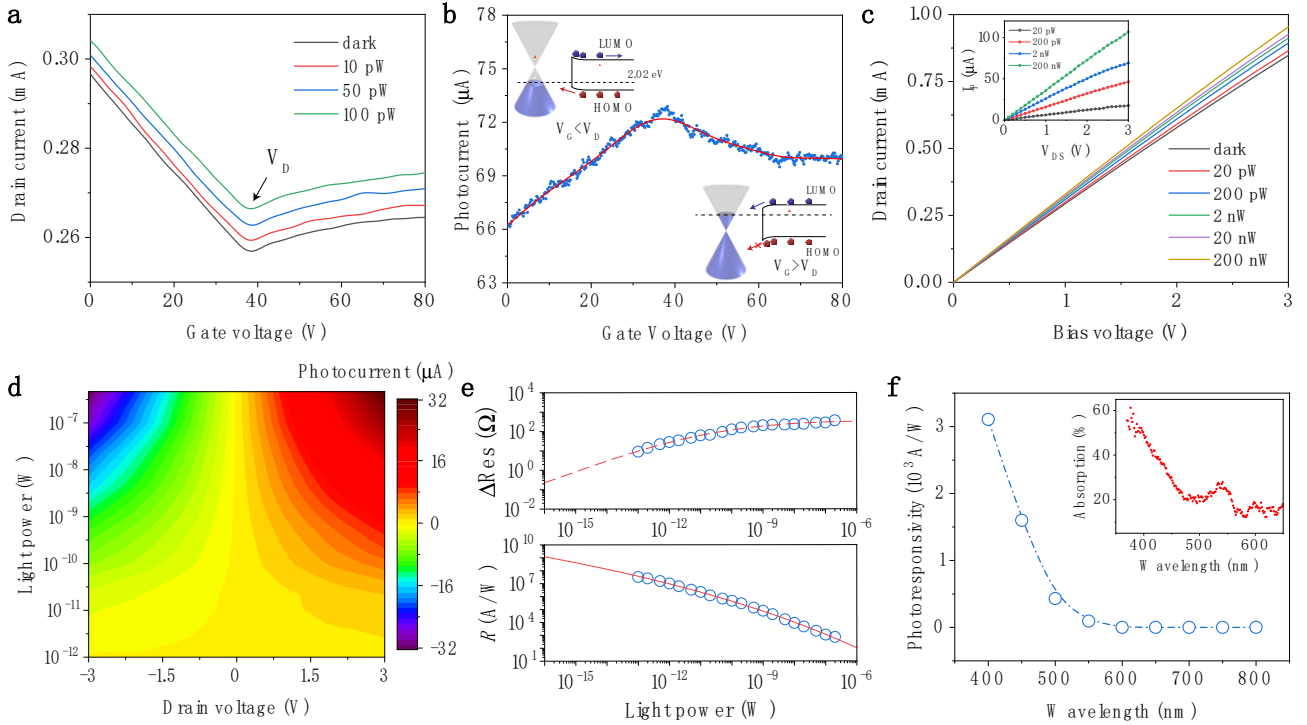

**Fig. 3. Photodetector device characteristics (measured at λ = 473 nm). (a)** Transfer curves ($V_{DS}$ = 1 V) of the photodetectors under different illumination powers. $V_D$ corresponds to the charge neutral point. **(b)** Photocurrent as a function of gate voltage $V_G$ under an illumination power of 100 nW. Inset: energy diagrams of the $COF_{ETBC-TAPT}$-graphene heterostructure. **(c)** Drain current as a function of bias voltage under different illumination powers at zero gate voltage. Inset: the calculated photocurrent as a function of the bias voltage. **(d)** Colormap of photocurrent generation relating to the illuminating power and bias voltage. **(e)** Light-induced resistance changes and photoresponsivity versus illuminating power ($V_G$ = 0, $V_{DS}$ = 3 V). The solid red line is the proper fitting of the measured data using the function $R=c_1+c_2/(c_3+P)$, where $c_1$, $c_2$ and $c_3$ are fitting parameters. **(f)** Photoresponsivity as a function of illumination wavelength from 400 to 800 nm. Inset: Optical absorption spectrum of the device.

The photoelectrical characteristics of the photodetectors are shown in Fig. 3. Fig. 3a shows the transfer characteristics (drain current, $I_{DS}$, vs. gate voltage, $V_G$) of the photodetector under different illumination powers from a 473-nm laser, with a fixed drain voltage ($V_{DS}$) of 1 V. The drain current of the photodetector consisted of a dark current (black line) and a photocurrent (defined as $I_p = I_{light} - I_{dark}$). The minimum drain-source current corresponding to the charge neutral point $V_D$ of the $COF_{ETBC-TAPT}$-graphene heterostructure, which suggests that the $COF_{ETBC-TAPT}$-graphene heterostructure is p-type doped and that the holes are majority carriers ($V_G$ = 0). High values of photocurrent were observed even at a very low illumination power (e.g., 1.79 μA at 2 pW), which are depicted in Fig. 3b. In the $V_G < V_D$ region, the carrier transport is hole dominated, and the photocurrent rises as the gate voltage increases. In the $V_G > V_D$ region, the $COF_{ETBC-TAPT}$-graphene heterostructure is electron doped, and the photocurrent slightly declines as the gate voltage increases. This can be explained by the energy diagrams in the inset of Fig. 3b. In this heterostructure, graphene provides a carrier transport channel with limited photoresponse, and COF$_{ETBC-TAPT}$ is used as the strong light-absorbing material. At the interface of COF$_{ETBC-TAPT}$ and graphene, a Schottky-like junction forms due to the injection of electrons from COF$_{ETBC-TAPT}$ into graphene. As a result, a built-in field with a direction from COF$_{ETBC-TAPT}$ to graphene is formed. In the V$_G$ < V$_D$ region, the energy band of COF$_{ETBC-TAPT}$ bends upwards at the interface of graphene. When the photodetector is under illumination, the COF$_{ETBC-TAPT}$ and graphene layer will generate electron-hole pairs. In the graphene layer, driven by the built-in field, the photoexcited electrons can move to the LUMO band of COF$_{ETBC-TAPT}$, while the photoexcited holes remain in graphene. In the COF$_{ETBC-TAPT}$ layer, photoexcited electrons are trapped due to the energy barrier, while photoexcited holes can be injected into the graphene layer. The trapped electrons in COF$_{ETBC-TAPT}$ serve as a negative local gate and thus induce a hole current in the graphene channel through capacitive coupling [39,40]. As a result, the recombination of photogenerated carriers can be suppressed efficiently, and the concentration of holes in the graphene layer increases, which results in a large positive photocurrent in the photodetector. Additionally, as the gate voltage increases, the Fermi energy of graphene increases to a higher level, which facilitates the injection of holes from COF$_{ETBC-TAPT}$ to the graphene channel, resulting in the rise of the photocurrent until V$_G$ = V$_D$. In the V$_G$ > V$_D$ region, graphene is shifted to electron doping, and the energy band of COF$_{ETBC-TAPT}$ is bending downwards at the interface of graphene. The injection of photoexcited electrons from the COF$_{ETBC-TAPT}$ layer to graphene is dominant in the heterostructure, while the photoexcited holes are trapped in the COF$_{ETBC-TAPT}$ layer. As the increasing gate voltage continues to increase the Fermi energy of graphene to a higher level, the built-in field between COF$_{ETBC-TAPT}$ and the graphene layer becomes weaker and finally leads to a slight decrease in the photocurrent. However, the increase in photocurrent in the V$_G$ < V$_D$ region is obvious, while the decrease in photocurrent in the V$_G$ > V$_D$ region is hardly visible in Fig. 3b, which may be explained by the difference in the rate of change in photoconductivity in these two regions.

Then, the zero gate voltage was applied to the photodetector (V$_G$ = 0 V), and the device turned into a standard photoconductor. The I$_{DS}$–V$_{DS}$ characteristic curves of the photodetector device under different illumination conditions are displayed in Fig. 3c. The linearity of the I$_{DS}$–V$_{DS}$ curve suggests an Ohmic contact between the Au electrodes and the COF$_{ETBC-TAPT}$-graphene heterostructure. The inset of Fig. 3c shows that the photocurrent is proportional to the drain voltages due to the increase in the carrier drift velocity. To further illustrate the relations between the photocurrent, the drain voltage and illumination power, the photocurrent mapping is depicted in Fig. 3d. The photoresponsivity ($R = I_p / P_{light}$) is an important parameter of a photodetector. The bottom panel of Fig. 3e shows the near-linear curve of photoresponsivity versus illumination power in a double logarithmic coordinate under zero gate voltage. We measured a photoresponsivity as high as ~3.2×10$^7$ A W$^{-1}$ for a light power of ~0.1 pW, followed by a decrease with increasing illumination intensity. The external quantum efficiency (EQE) is a parameter related to photoresponsivity, which is calculated by $EQE = R(hc/e\lambda)$, where $h$ is Planck's constant, $c$ is light speed in vacuum, $e$ is the quantity of electric charge, and $\lambda$ is the light wavelength. The corresponding EQE reaches ~$8.5 \times 10^9$ % under the same conditions. According to the calculated noise level floor of 0.24 $\Omega \cdot Hz^{-1/2}$ (Fig. S9), which determines the ultimate sensitivity of the photodetector (top panel of Fig. 3e), the corresponding noise-equivalent power (NEP) is ~$10^{-16}$ W. The photodetectivity ($D^* = (A\Delta f)^{1/2} R / i_n$) of the devices was calculated under the same condition and reaches up to ~$6 \times 10^{13}$ Jones (cm $Hz^{1/2}$ $W^{-1}$) [41]. The $D^*$ and NEP values are comparable to those of commercial silicon photodiodes and are mainly limited by the relatively large dark current of graphene.[42] To investigate the detection spectrum of the device, the photoresponsivity as a function of the illumination wavelength is displayed in Fig. 3f. The device shows decreasing photoresponsivity as the illumination wavelength increases from 400 nm to 800 nm and a cut-off at 600 nm, which is consistent with the absorption curve of the device in the inset of Fig. 3f and the measured energy bandgap of $COF_{ETBC-TAPT}$ in Fig. S5.

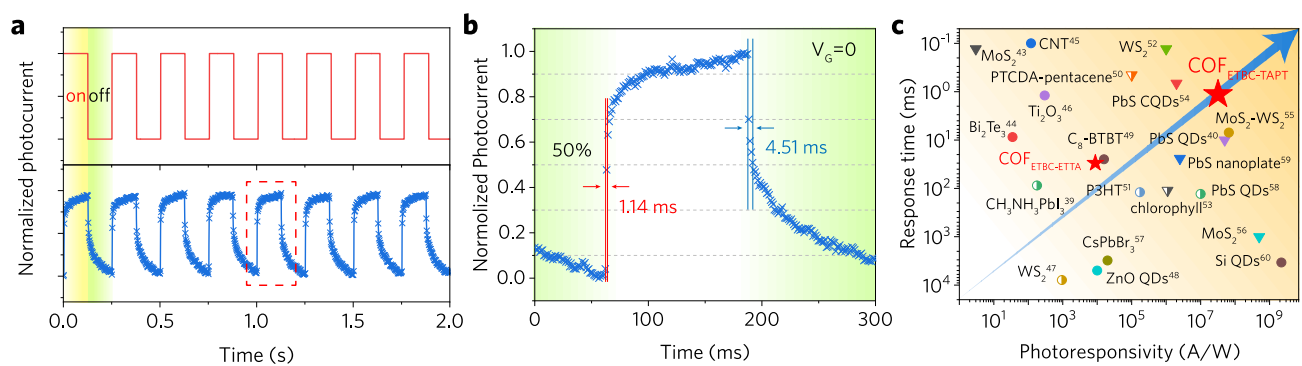

**Fig. 4. Photocurrent dynamics of the device.** (a) Photoswitching performance under alternating dark and light illumination ($V_G = 0$, $V_{DS} = 1$ V, $\lambda = 473$ nm). (b) An enlarged view of the normalized photocurrent dynamics during one cycle of light modulation. (c) Summary of the device performance of graphene-based photodetectors with different semiconductors. The circles are devices measured at zero gate voltage. The triangles are devices using a vertical field. The half-empty points are devices using fitting methods to calculate the response time.
$MoS_2$[43] $Bi_2Te_3$[44] CNT[45] Perovskite ($CH_3NH_3PbI_3$) [39] $Ti_2O_3$[46] $WS_2$[47] ZnO QDs[48] $C_8$-BTBT[49] GPP[50] P3HT[51] $WS_2$[52] chlorophyll[53] PbS CQDs[54] PbS QDs[40] $MoS_2$-$WS_2$[55] $MoS_2$[56] $CsPbBr_3$[57] PbS QDs[58] PbS nanoplate[59] Si QDs[60]

To confirm the temporal photoresponse characteristic of $COF_{ETBC-TAPT}$-graphene photodetectors, the normalized photocurrent with periodically switched illumination was measured under a bias voltage of 1 V, as shown in Fig. 4a. The photodetector exhibited stable on-off switching synchronized with illumination. We repeated the on-off cycles of illumination over 800 times, showing that the photodetector exhibits great stability (Fig. S10a). The rise/fall times corresponding to 3 dB lower than the signal peak were measured to be ~1.14 ms and ~4.51 ms, as shown in Fig. 4b. The 0%-80% rise/fall time was estimated to be ~6.81 ms and ~46.35 ms, respectively (Fig. S10b). In the AB-stacked $COF_{ETBC-TAPT}$ structure, there is no obvious continuous pathways for carrier transport along the stacking direction, which might be a reason for the relatively slow response times.

The performance of state-of-the-art photodetectors interfacing graphene with different photoactive materials is summarized in Fig. 4c. The direction of the arrow in Fig. 4c represents the trend of optimization and idealization for the photodetector devices. In this study, by synthesizing a variety of 2D-COFs, we found that the 2D-COFs synthesized with the selection of TPE monomers tended to obtain optoelectronic properties, such as $COF_{ETBC-ETTA}$, synthesized by selecting ETBC to react with 4,4′,4″,4‴-(ethene-1,1,2,2-tetrayl)-tetraaniline (ETTA) (see Supplementary Information for characterization details). However, constructed by two types of TPE monomers, $COF_{ETBC-ETTA}$ possesses no electron acceptors, which limits its performance. Thus, we chose a triazine monomer (TAPT) to replace one of the TPE monomers (ETTA) and synthesized $COF_{ETBC-TAPT}$, constituting an electron donor-acceptor structure to facilitate charge separation and electron transfer. In $COF_{ETBC-TAPT}$, TPE motifs (ETBC) can act as moderated electron donors, while triazine motifs (TAPT) are strong electron acceptors. Considering the trade-off between photoresponsivity and time response, the results and measuring conditions of previous works are further summarized in Table S1. By comparison, the $COF_{ETBC-TAPT}$-graphene photodetector displays an excellent overall performance, showing high potential to be further optimized by modifying the monomers, which reflects that COF is a good platform for the preparation of functional optoelectronic devices with broader application prospects.

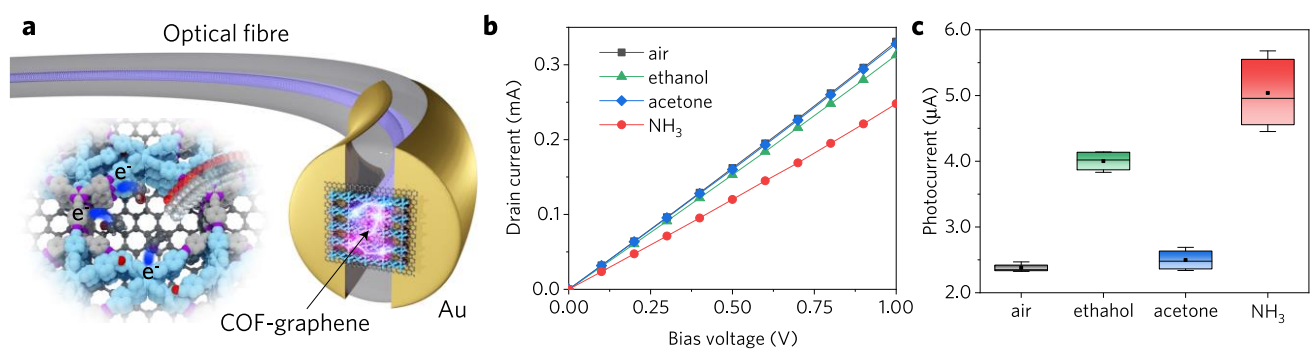

**Fig. 5. Optoelectronic properties regulated by target molecules.** (a) Schematic view of the gas molecule absorption and charge transfer at the surface of the $COF_{ETBC-TAPT}$-graphene film. (b) $I_{DS}$–$V_{DS}$ characteristic curves in the dark with different gas molecules (air, 1% ethanol vapor in air, 1% acetone vapor in air and 1% $NH_3$ in air). (c) Photocurrent generation in different gas atmospheres.

In addition to optimizing the combination of π-electron monomers, COF-graphene photodetectors can be further regulated by the external environment[61,62]. COFs have been proven to be ideal materials for gas adsorption due to their highly porous structure with high surface areas. Here, Nitrogen sorption analysis was carried out to measure the porosities of $COF_{ETBC-TAPT}$ (Fig. S3). The $COF_{ETBC-TAPT}$ possesses high Brunauer-Emmett-Teller (BET) surface area of 1745 $cm^2\ g^{-1}$(Fig. S4). The main pore-size distributions of $COF_{ETBC-TAPT}$ calculated by nonlocal density functional theory (NLDFT) are at 1.4 nm, which is in good agreement with the proposed structures.

Thus, we investigated the gas effects on the photoelectrical properties of the COF$_{ETBC-TAPT}$-graphene photodetector with different gas molecules. Here, considering the small size and flexibility of the optic fibre platform, which is more conducive to the detection of the external environment, an optical-fibre-compatible photodetector (FPD) was successfully demonstrated by integrating the COF$_{ETBC-TAPT}$-graphene heterostructure film on the optical fibre end-face (Fig. S7). From the I$_{DS}$–V$_{DS}$ curves in the dark in Fig. 5b and the enhanced photocurrent generation in Fig. 5c at various gas atmospheres, the photodetector can strongly respond to the given gas molecules—especially strongly polar molecules like NH$_3$ and ethanol—which play important roles in the regulation of the photoelectric performance of the device. This response is considered to result from the charge transfer between the COF$_{ETBC-TAPT}$-graphene heterostructure and the adsorbed gas molecules, as shown in Fig. 5a. Once the gas molecules come into contact with COF$_{ETBC-TAPT}$, they will be adsorbed and subsequently transfer electrons and change the charge carrier distribution in the heterostructure, affecting the photoelectric characteristics of devices [63,64]. The COF$_{ETBC-TAPT}$ contain aldehyde groups in the pores, which result in strong adsorption for NH$_3$ and other polar gas molecules. Therefore, these polar gas molecules have a greater influence on the photoelectric performance of the device. The gas sensitivity of a device is defined as $S = (R_{gas} - R_{air})/R_{air}$, where $R_{gas}$ is the resistance of the device in the target gas and $R_{air}$ is the resistance of the device in air. Fig. S11 shows one-cycle absorption and desorption responses for 1% NH$_3$ in air and 1% ethanol vapor in air; the gas sensitivities are measured to be 16.2% and 5.8%, respectively. The response time of gas absorption and desorption are slow (in minutes scale), which might because that the pores of COF were already filled with gas molecules in the air environment, so the adsorption and desorption of specific gas molecules had a process of gas molecule replacement and diffusion. Moreover, the COF$_{ETBC-TAPT}$ structure is AB-stacked without continuous pathways for gas molecules transport, which will also affect the time response of gas sensing. Due to the porous nature of COF materials, the optoelectronic properties of the photodetectors can be reversibly regulated by target gas molecules, indicating potential applications in sensor fields.

In summary, we propose a strategy to synthesize photosensing 2D-COFs by selecting suitable monomers with photoelectric activities. Well-ordered COF-graphene heterostructures were prepared by an in situ growth process. Ultrasensitive photodetectors with excellent overall performance were successfully fabricated and demonstrated. Moreover, owing to the high surface area and the polarity selectivity of COFs, the photodetectors can be strongly regulated by specific target molecules. Flexible structure design and external regulation of COFs will open a route towards achieving advanced optoelectronics and many other applications.

## Methods

**Synthesized COF$_{ETBC-TAPT}$-graphene heterostructure.** The CVD-grown graphene supported on a Cu substrate was put into a glass tube containing 44.9 mg of 4',4''',4'''',4'''''-(1,2-ethenediylidene)tetrakis[1,1'-biphenyl]-4-carboxaldehyde (ETBC, 98%), 28.3 mg of 2,4,6-tris(4-

aminophenyl)-1,3,5-triazine (TAPT, 98%) and 0.45 mL of mixed solvent of o-dichlorobenzene/n-butanol/12 M acetic acid (v/v, 48:12:5). The glass tube was then flame-sealed and heated at 120 ℃ for 1 day. After cooling to room temperature, the mixture was rinsed with THF at least 5 times, purified by Soxhlet extraction with THF for 24 hours, and dried under supercritical $CO_2$ flow for 3 h.

**Device fabrication.** The sequential fabrication process of the silicon-based photodetectors: First, an array of Au electrodes (60 nm thick, 10 μm length of gap, made by photolithography) were magnetron sputtered on a silicon (Si) substrate with a 285-nm-thick silicon dioxide ($SiO_2$) layer, and then, the prepared $COF_{ETBC-TAPT}$-graphene thin film was transferred to it via the wet transfer method. Photolithography and $O_2$ plasma etching were carried out to remove redundant areas of $COF_{ETBC-TAPT}$-graphene film and pattern the channel.

The sequential fabrication process of the optical-fibre-compatible photodetectors: First, the Au layer (40 nm thick) was magnetron sputtered on a cleaved single-mode optical fibre. A focused ion beam (FIB) process was used to obtain an electrode channel with a 10 μm gap length on the optical fibre end-face. A portion of the gold layer at the lateral wall of the fibre was scratched to obtain a small electrode channel. Then, the prepared COF-graphene film was transferred onto the electrodes on the optical fibre end-face and placed to cover its core by a 3D micro-operating transfer method (see supplementary information for details).

**Measurement equipment.** All silicon-based measurements were tested by a probe station (Cascade Summit 12000B-M) in a dark environment. The characteristics of the device were collected and analysed by a parameter analyser (Keithley 4200A-SCS, Tektronix). Additionally, an optical fibre was used to transmit the light illumination to the device, and a reference optical power-metre (S150C and S145C, Thorlabs) was used to calibrate the input light power. In the photocurrent dynamics test, an optical chopper was used (Model C-995, Scitec) to realize the illumination ON-OFF switches.

The optical-fibre-compatible photodetectors (FPDs) were measured in a dark sealed box with a gas channel. Since the FPD was naturally compatible with optical fibre systems, the incident light was transported by the optical fibre waveguide and used to illuminate the FPD directly. A reference optical power-metre was also used (S150C and S145C, Thorlabs) to calibrate the input light power. The electrical signal was collected and analysed by a digital source-metre (Keithley SMU 2450, Tektronix).

# Acknowledgments

We thank Prof. W.H. Zhang, Prof. X.J. Zhang and Mr. K. Tai for help in the measurement of Ramam spectra and AFM images of $COF_{ETBC-TAPT}$-graphene samples. This work was sponsored by the National Key R&D Program of China (2017YFA0303700 and 2017YFA0700503), the National Natural Science Foundation of China (61535005), the Fundamental Research Funds for the Central Universities (020514380190) and the Scientific Research Foundation of the Graduate School of Nanjing University (2018CL02).

## Author contributions

F.X. and K.X. conceived the experiments. Y.F.X. and H.T.Z fabricated and measured the photodetectors. Q.B.L., X.H. and C.K. synthesized the 2D-COFs and contributed to material characterization, Q.B.L. synthesized $COF_{ETBC-TAPT}$-graphene heterostructure. Y.F.X., Q.B.L., F.X. and K.X. analyzed the data. C.Y.D and H.S.W. contributed to material and device characterization. Z.P.H. and P. Z. analyzed Raman spectroscopy. Y.Q.L. advised on the experiments. Y.F.X. wrote the original draft. F.X., P.Z., K.X. and Y.Q.L. reviewed and edited the paper. F.X. guided the research and supervised the project.

## Competing interests

The authors declare that they have no competing interests.